\documentclass[12pt]{iopart}
\usepackage[utf8]{inputenc}
\usepackage[english]{babel}
\usepackage{graphicx}
\usepackage{hyperref}

\usepackage{braket}

\usepackage{xspace}
\newcommand{\BePlus}{$^9$Be$^+$\xspace}

\newcommand{\pbars}{\mbox{(anti-)}protons\xspace}

\usepackage{float}

\begin{document}

\title[139\,GHz UV phase-locked Raman laser system]{139\,GHz UV phase-locked Raman laser system for thermometry and sideband cooling of \BePlus ions in a Penning trap}

\author{J Mielke$^1$, J Pick$^1$, J A Coenders$^1$, T Meiners$^1$, M Niemann$^1$, J M Cornejo$^1$, S Ulmer$^2$, C Ospelkaus$^{1,3}$}

\address{$^1$ Institut für Quantenoptik, Leibniz Universität Hannover, Welfengarten 1, 30167 Hannover, Germany}
\address{$^2$ RIKEN, Ulmer Fundamental Symmetries Laboratory, 2-1 Hirosawa, Wako, Saitama 351-0198, Japan}
\address{$^3$ Physikalisch-Technische Bundesanstalt, Bundesallee 100, 38116 Braunschweig, Germany}
\ead{christian.ospelkaus@iqo.uni-hannover.de}

\begin{abstract}
We demonstrate phase locking of two ultraviolet laser sources by modulating a fundamental infrared laser with 4th-order sidebands using an electro-optic modulator and phase locking of one sideband to a second fundamental infrared laser. Subsequent sum frequency generation and second harmonic generation successfully translates the frequency offset to the ultraviolet domain. The phase lock at 139\,GHz is confirmed through stimulated Raman transitions for thermometry of \BePlus ions confined in a cryogenic Penning trap. This technique might be used for sideband cooling of single \BePlus ions as well as sympathetic cooling schemes and quantum logic based measurements in Penning traps in the future.
\end{abstract}
\submitto{\jpb}

\maketitle

\section{Introduction}

High precision measurements of the properties of fundamental particles like the free \cite{hanneke_new_2008} or bound electron \cite{sturm_high-precision_2014, arapoglou_g_2019}, the proton \cite{schneider_double-trap_2017} or the antiproton \cite{ulmer_high-precision_2015, smorra_350-fold_2018} provide stringent tests of the foundations of the Standard Model such as quantum electrodynamics (QED) and charge-parity-time (CPT) symmetry \cite{particle_data_group_review_2018, lehnert_cpt_2016}. Important examples are charge-to-mass ratio or magnetic moment measurements in Penning traps, which are based on cryogenic tank-circuits for image-current detection of the trapped particle's eigenfrequencies \cite{wineland_principles_1975}. Present experiments fight against systematic uncertainties depending on the motional amplitudes of the particle  \cite{brown_geonium_1986} and efforts are currently under way to implement sympathetic laser cooling schemes in order to further reduce the motional amplitude \cite{cornejo_optimized_2016, bohman_sympathetic_2018, sturm_alphatrap_2019} or aim for cooling the particle to the quantum mechanical ground state of motion \cite{meiners_towards_2018}. Moreover, new quantum logic schemes for ground state cooled particles in Penning traps have been proposed in order to perform faster measurements with higher fidelities and statistics \cite{wineland_experimental_1998-1, cornejo_quantum_2021}.

For sympathetic cooling of single particles it is desirable to choose a small mass-ratio for the particle of interest and the cooling ion in order to maximize the efficiency of the cooling process \cite{wubbena_sympathetic_2012}. Thus, as the lightest readily laser-cooled atomic ion species, \BePlus is the natural choice for sympathetic cooling of \pbars. Furthermore, sympathetic cooling schemes and quantum logic spectroscopy are most readily implemented on the axial motion in Penning traps  due to the relative ease of motional ground state preparation. In this case optimal performance is achieved when the interaction with the radial motion is minimized. Using \BePlus this  situation can be realized with two phase-locked lasers for induced Raman transitions between the sublevels of the $1s^22s$ ground state level if the lasers are aligned so that their effective wavevector difference is oriented along the axial direction.

Due to the experimental complexity of the tools and techniques required for sympathetic cooling and quantum logic spectroscopy of \pbars, it is useful to verify their functionality in simplified proof-of-principle experiments. For example, in order to prove the viability of the laser setup that is required for ground state cooling of single \BePlus ions, we can also perform axial thermometry measurements with \BePlus clouds, which are easier to prepare and to control than single ions. By probing the Raman resonance of the ion cloud it is then possible to calculate the effective mode temperature from the equivalent Doppler width of the resonance, similar to the case of single ions \cite{mavadia_optical_2014}.

\BePlus has optical resonances at 313\,nm and the characteristic high magnetic field of a Penning trap introduces large Zeeman shifts on the order of hundreds of GHz. Thermometry and sideband cooling based on Raman transitions therefore might be implemented using a tailored ultraviolet (UV) frequency comb \cite{paschke_versatile_2019}. However, beam delivery without UV-compatible fibers \cite{colombe_single-mode_2014, marciniak_towards_2017} suitable for usage with pulsed lasers as well as time synchronization of laser pulses with ps pulse lengths  in a trap with limited optical access makes it more challenging to integrate such a system in an existing experiment. On the other hand, the lack of UV-compatible high frequency modulators and the need for high-speed detection electronics complicates an implementation with continuous wave (cw) lasers as well.

Here we introduce a method for phase locking of two cw laser sources at 313\,nm that utilizes higher order sideband modulation and phase locking of two fundamental lasers at 1552\,nm in order to generate the offset frequency required for stimulated Raman transitions with  \BePlus ions in a high magnetic field. We use the nonlinear conversion processes inherent to our laser system to further increase the offset frequency and to transfer the phase lock to the UV domain. We demonstrate the usability of this technique with an axial thermometry measurement of \BePlus ions confined in a cryogenic Penning trap.

\section{Laser setup and characterization}

\begin{figure}[b]
	\centering
	\includegraphics[width=1\columnwidth]{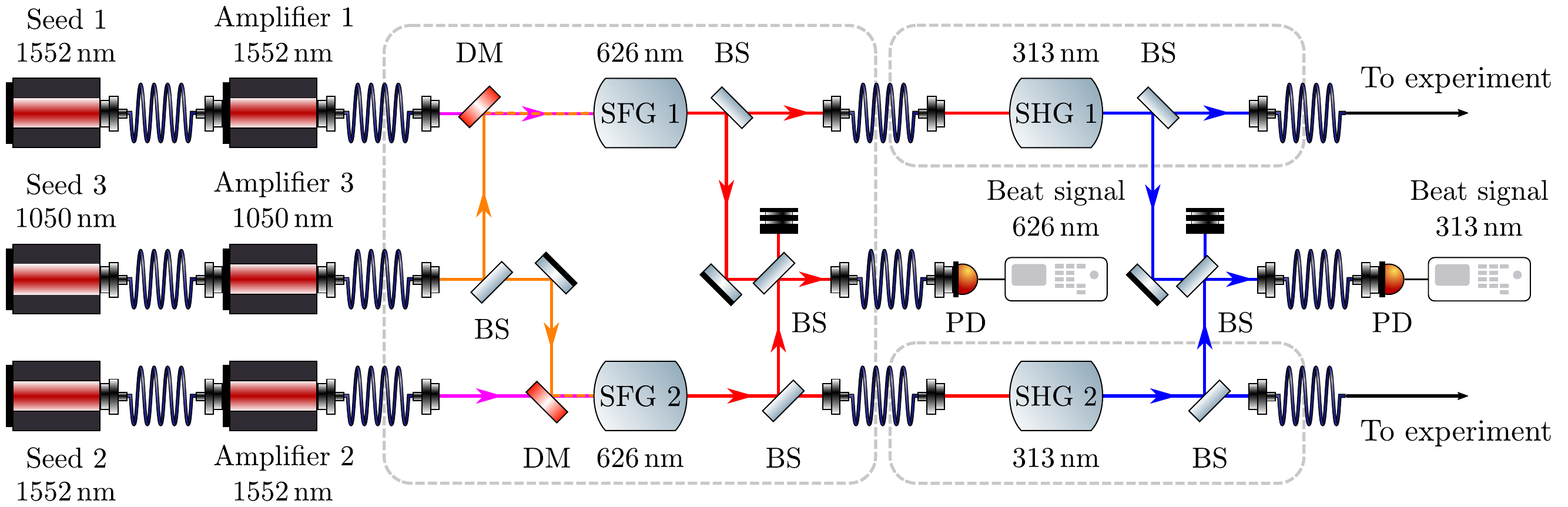}
	\caption{Schematic of the laser system for generation of two independent laser beams at 313\,nm. The setup is divided into subsystems for laser amplification, SFG and SHG. Grey dashed lines indicate enclosed laser boxes which are connected via fiber links. Beat signals are measured after the nonlinear frequency conversion processes at the labeled positions. BS: beam splitter, DM: dichroic mirror, PD: photodiode.}
	\label{fig:lasersystem}
\end{figure}

Raman transitions for thermometry are implemented between the sublevels $^2S_{1/2}\ket{m_I=+3/2, m_J=+1/2}$ and $^2S_{1/2}\ket{m_I=+3/2, m_J=-1/2}$ of the \BePlus  ground state, with a  Zeeman splitting of 139\,GHz at a magnetic field strength of 5\,T \cite{paschke_9be+_2017}. The coupling between the states is faciliated by off-resonant interaction with the $1s^22p$ excited state manifold using two lasers at 313\,nm. The setup for generation of the laser beams is depicted in Fig.~\ref{fig:lasersystem}. It is based on two amplified fiber lasers near 1552\,nm (\textit{NKT Photonics, Koheras BOOSTIK HP E15}) and a third one at 1050\,nm (\textit{NKT Photonics, Koheras BOOSTIK HP Y10}), whose output is first converted into two laser beams at 626\,nm using sum frequency generation (SFG) and subsequently  to 313\,nm using second harmonic generation (SHG) \cite{wilson_750-mw_2011}. As we follow a modular approach, the two SFG setups are housed in a single enclosed laser box, while each SHG stage is built as a separate module. The fundamental lasers and the boxes are connected via fibers, which enables for a flexible placement in limited lab spaces. The 313\,nm light is guided to our experiment via solarization-resistant photonic crystal fibers suited for UV transmission \cite{colombe_single-mode_2014, marciniak_towards_2017}. Each frequency doubling cavity of the SHG stages is able to provide a power of 160\,mW of UV light of which only a few mW are used in the experiment.

As it is hard to generate and detect the required offset frequency with modulators and photodetectors at the output UV frequencies, we rather implement the phase lock using the infrared (IR) sources, similar to Ref. \cite{jordan_near_2019}. This allows us to use fast fiber-based IR modulators and photodetectors in the signal chain of the phase lock. In addition we benefit from the fact that frequency doubling of the fundamental lasers will double the frequency offset as well.

\begin{figure}[t]
	\centering
	\includegraphics[width=0.5\columnwidth]{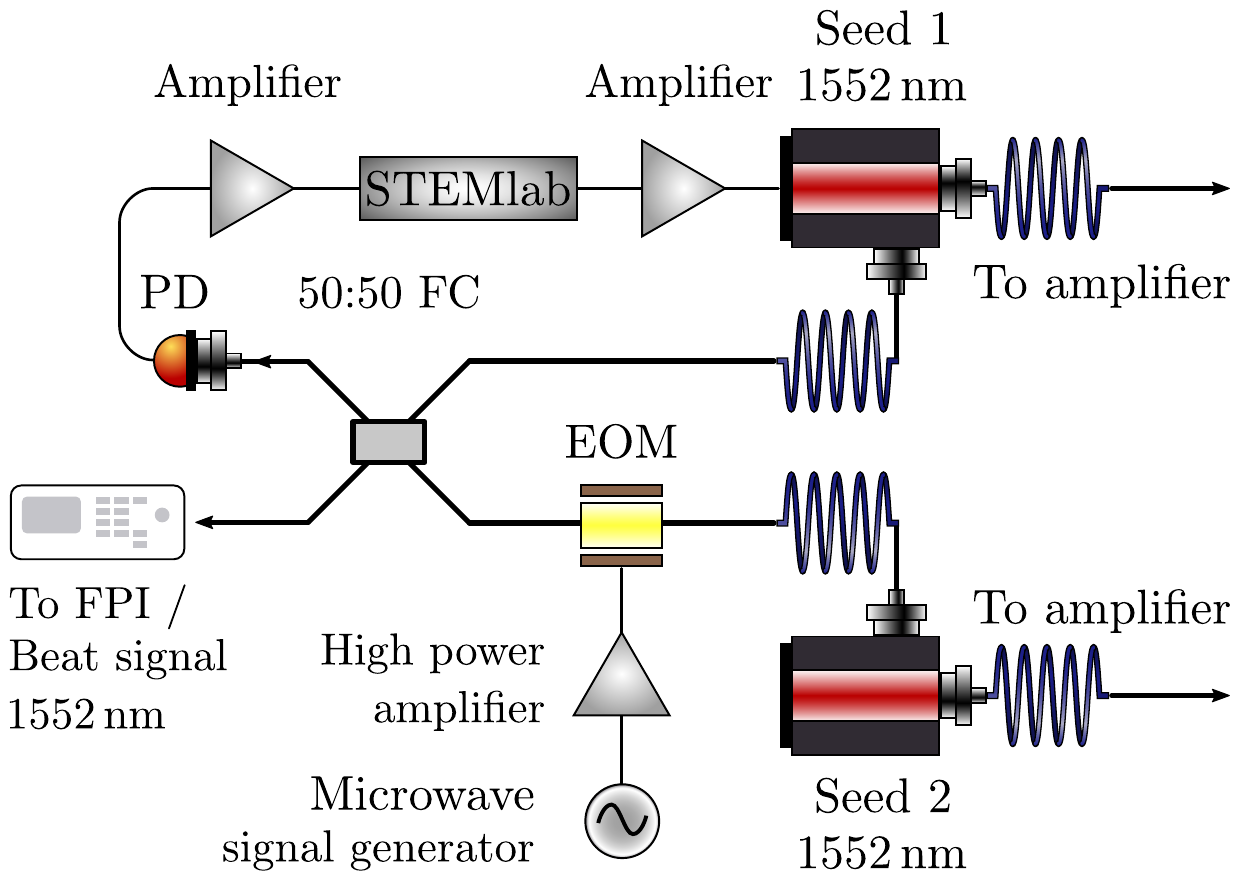}
	\caption{Schematic of the servo loop for the phase lock of the seed fiber lasers at 1552\,nm. FC: fiber coupler, FPI: Fabry-Pérot interferometer, PD: photodiode.}
	\label{fig:PLL}
\end{figure}

We ensure the frequency offset before SHG through a phase lock of the two fundamental seed lasers at 1552\,nm with the setup shown in Fig.~\ref{fig:PLL}. The reference output of seed laser 2 is modulated with a fiber-coupled electro-optic modulator (EOM, \textit{EOSPACE, PM-AV5-40-PFA-PFA-SRF1W}) at a frequency of  17.375\,GHz. The modulation signal is generated by a microwave signal generator (\textit{Anritsu, MG3692B} with option for ultra low phase noise) and amplified with a high power amplifier (\textit{Mini-Circuits, ZVE-3W-183+}) up to 30\,dBm, which is the maximum rated input power of the EOM. This setup allows us to generate and maximize the 4th-order sideband component at 69.5\,GHz. A combination of a higher modulation frequency and lower sideband order could be used as well, but was dismissed in this case, as suitable microwave sources and amplifiers would significantly increase the overall cost of the setup without reasonable improvement of the performance. Seed laser 1 is tuned close to this sideband and its reference output is overlapped with the modulated laser using a 2x2 fiber coupler (\textit{Agiltron, FC-15139233}). The resulting beat signal is measured with a fast fiber-coupled photodetector (\textit{Menlo Systems, FPD310-FC-NIR}). Although the beat signal frequency is usually kept at 31.25\,MHz, the high bandwidth of the detector is helpful for finding the signal initially. After amplification the beat signal is fed into a \textit{Red Pitaya STEMlab 125-14}  board that implements a digital phase-locked loop (PLL) \cite{tourigny-plante_open_2018}. The beat signal is thereby compared and phase-locked to a fractional frequency of the board's internal 125\,MHz quartz oscillator. The output of the digital controller is amplified and fed back to the frequency modulation input of seed laser 1, thus closing the feedback loop of the phase lock.

\begin{figure}[b]
	\centering
	\includegraphics[width=0.5\columnwidth]{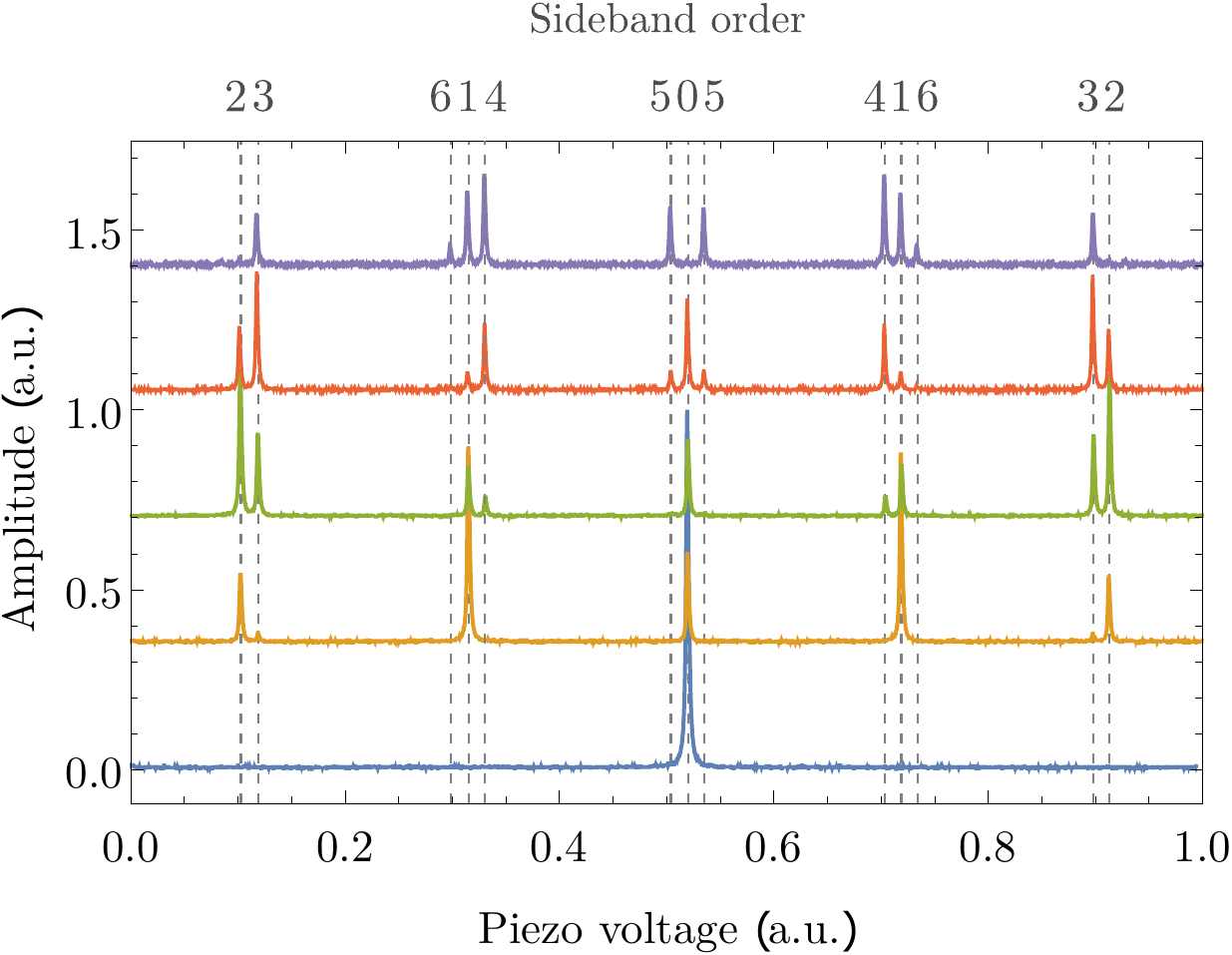}
	\caption{Sideband spectra at 1552\,nm measured with a Fabry-Pérot interferometer. The plot displays a resonance scan with modulation turned off (blue) and four resonance scans at EOM driving power levels of 19.0\,dBm (yellow), 24.0\,dBm (green), 26.2\,dBm (red) and 28.0\,dBm (purple). The power levels correspond to modulation indices that maximize the 1st-, 2nd-, 3rd- and 4th-order sideband, respectively. The amplitude scale is identical for all measurements, while the individual measurements are offset by a constant value for clarity. Grey dashed lines and labels on top indicate sideband positions and sideband orders.}
	\label{fig:sidebandspectra}
\end{figure}

For a measurement of the sideband spectrum, we turn on seed laser 2 only and couple the beam from the fiber coupler's second output port into a Fabry-Pérot interferometer (FPI, \textit{Thorlabs, SA200-12B}). We scan the length of the interferometer by applying a triangular voltage to the piezoelectric transducer, while monitoring the response of the integrated photodiode detector with an oscilloscope. The recorded spectra are shown in Fig.~\ref{fig:sidebandspectra}. Without modulation only a single resonance peak is visible. As we turn on the modulation and increase the driving power more sidebands appear in the spectrum. Since the interferometer has a free spectral range of 1.5\,GHz, the sidebands appear at positions corresponding to the modulus of their respective frequency. The amplitudes of the sidebands depend on the EOM driving power and the relative signal strength can be deduced from a Bessel function expansion of the modulation signal \cite{peng_locking_2014}. In our case the amplitudes of the 1st-, 2nd-, 3rd- and 4th-order sidebands are maximized for EOM driving powers  of 19.0\,dBm, 24.0\,dBm, 26.2\,dBm, 28.0\,dBm, respectively. We also observe higher order sidebands, but are unable to maximize them, as the driving power is limited by the damage threshold of our EOM.

For phase locking, we set the EOM driving power to 28.0\,dBm in order to maximize the signal-to-noise ratio (SNR) of the beat signal for the digital PLL. Phase locking of seed laser 1 to the 4th-order sideband of seed laser 2 thus provides us with a sub-GHz frequency of the beat signal. As we only modulate the reference output of seed laser 2, no sidebands at 17.375\,GHz or multiples thereof are present in the main amplified laser beam, maintaining all the power in the carrier and avoiding spurious transitions.

As we lack the possibility to measure and analyze the beat signal at 69.5\,GHz directly, we measure the beat signal with a reduced offset frequency at different positions in the laser system instead. To do so, we turn off the microwave modulation and lock seed lasers 1 and 2 to each other with a frequency offset of  only 31.25\,MHz. Since the additional phase noise introduced by the microwave signal is negligible even for the 4th-order sideband component (the specified single-sideband phase noise at 17.375\,GHz is -45\,dBc/Hz at an offset frequency of 10\,Hz and -94\,dBc/Hz at an offset frequency of 1\,kHz), this allows for an analysis of the degradation of the phase lock throughout the setup and findings can be applied to the case with high offset frequency as well.

\begin{figure}[b]
	\centering
	\includegraphics[width=1\columnwidth]{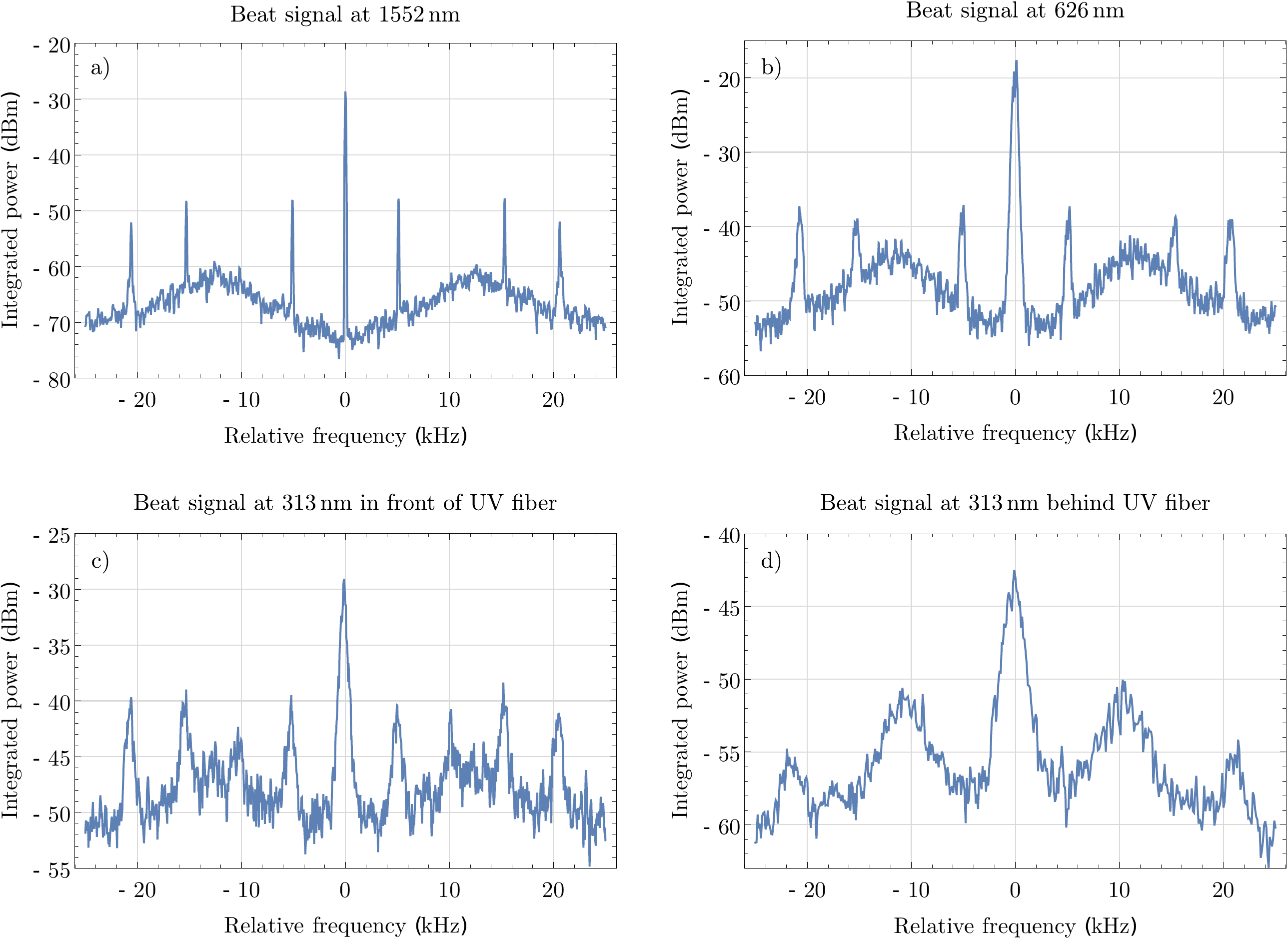}
	\caption{Beat signals spectra measured at 1552\,nm, 626\,nm, 313\,nm. The frequencies are given relative to a  central frequency of 31.25\,MHz for the measurements at 1552\,nm, 626\,nm and relative to a central frequency of 62.5\,MHz for the measurements at 313\,nm. All spectra correspond to an average of 10 scans with a 100\,Hz measurement bandwidth each.}
	\label{fig:beat_signals}
\end{figure}

For the analysis, we measure the beat signal spectrum at four different positions in the setup and compare the changes of the main spectral features. The first spectrum  is measured at 1552\,nm with a fiber-based photodiode connected to the second output port of the fiber coupler that is used for overlapping the reference output of the seed lasers (see Fig.~\ref{fig:PLL}). The beat signal at this position is comparable to a signal that would be obtained with the main output of both seed lasers before injecting them into the fiber laser amplifiers. The second spectrum is obtained at 626\,nm just after SFG by overlapping both lasers and sending them to a fiber-coupled photodiode (see Fig.~\ref{fig:lasersystem}). The third spectrum is recorded at 313\,nm in similar manner just after SHG. The fourth spectrum is measured after transmission of the 313\,nm lasers through the UV fibers. The two beams are picked off from the main beams and focused onto a photodiode with a lens (see Fig.~\ref{fig:experimentalsetup}). All resulting beat signal spectra are recorded with a spectrum analyzer and shown in Fig.~\ref{fig:beat_signals}.

At 1552\,nm a central peak with a linewidth smaller than the minimal bandwidth of the spectrum analyzer ($<1\,\mathrm{Hz}$) is observed in the beat signal spectrum together with broad bumps around a relative frequency of 12.6\,kHz (Fig.~\ref{fig:beat_signals}a). The central frequency of these ``servo bumps'' corresponds to the frequency where the phase delay of the feedback signal exceeds $\pi$, thus marking the transition from negative to positive feedback. Additional narrow parasitic sidebands appear in the spectrum at 5.0\,kHz, 15.3\,kHz, 20.6\,kHz, which we attribute to the seed lasers' internal power switching supplies, as they are reproducible with all lasers of the same type that we own. We measure a SNR of 19.5\,dB (where we refer to the difference between the main peak and the strongest parasitic sideband).

After the fiber-based amplification step at 1552\,nm and subsequent conversion to 626\,nm, a broadened central peak with a full width at half maximum (FWHM) of about 500\,Hz is observed in the beat signal. The broadening is introduced as the fibers couple the laser frequencies to the laboratory's thermal and acoustic environment \cite{williams_high-stability_2008}. The SFG process itself does not introduce additional phase noise, as the same 1050\,nm laser is used in both setups, where it follows a similar optical path. A similar amount of broadening is observed for the parasitic sidebands as well as an overall increase of the background noise (including the servo bumps). Since the amplitudes of the parasitic sidebands are affected in the same way as the amplitude of the main peak, the SNR is left unaffected (Fig.~\ref{fig:beat_signals}b).

Additional broadening is introduced by the following fiber links at 626\,nm, each 5\,m long. In addition, the SHG process leads to a doubling of the central beat frequency to 62.5\,MHz as well as a doubling of the width of all components in the spectrum. We measure a FWHM of about 1\,kHz of the central peak. Furthermore, the frequency doubling introduces additional parasitic sidebands at mixing frequencies of the already existing ones, with the most prominent one at 10.3\,kHz. The SNR is reduced to 9.3\,dB by this process (Fig.~\ref{fig:beat_signals}c).

In the final spectrum, measured after transmission of the laser beams at 313\,nm through the 6\,m UV fibers, the central beat signal peak is broadened to a FWHM of 1.5\,kHz. An additional broad noise peak at 10.4\,kHz is observed, which buries the parasitic sidebands. We attribute this effect to a decreased signal strength because of the free space overlap of the laser beams in combination with additional thermal and acoustic noise that is introduced by the fibers  guided close to the main experiment, where acoustic noise sources such as turbo molecular pumps are located. We measure a SNR of 7.6\,dB at this position (Fig.~\ref{fig:beat_signals}d). Within a frequency span of $\pm100\,\mathrm{kHz}$ ($\pm500\,\mathrm{kHz}$), the background noise drops by 32\,dB (64\,dB) in comparison with the main peak due to the intrinsic low phase noise of the seed fiber lasers at 1552\,nm.

This needs to be compared with the spectral requirements for axial thermometry of an ion crystal and for sideband cooling of single ions. For the former, we need to resolve Doppler broadened resonances on the order of a few MHz. For the latter, we have tighter requirements as typical axial trap frequencies in Penning traps are on the order of hundreds of kHz and individual sidebands needs to be resolved. With a linewidth on the order of a kHz at 313\,nm, both requirements are fulfilled by the phase-locked laser system. 

 The measurements show that the width of the beat signal could be reduced by guiding the fibers in more rigid and isolated enclosures or by using techniques for active fiber noise cancellation. On the one hand, phase noise introduced through the fiber-based amplification step at 1552\,nm can be mitigated by implementing the phase lock after amplification. On the other hand, noise introduced by the fiber links at 626\,nm and 313\,nm can be suppressed by using a double-pass fiber noise cancellation setup \cite{ma_delivering_1994}. However, in order to keep the experimental overhead as small as possible and to avoid modification of existing laser designs with limited space margins, we decided to omit the implementation of these techniques in the setup described here.
 
\section{Experimental verification}

\begin{figure}[t]
	\centering
	\includegraphics[width=0.6\columnwidth]{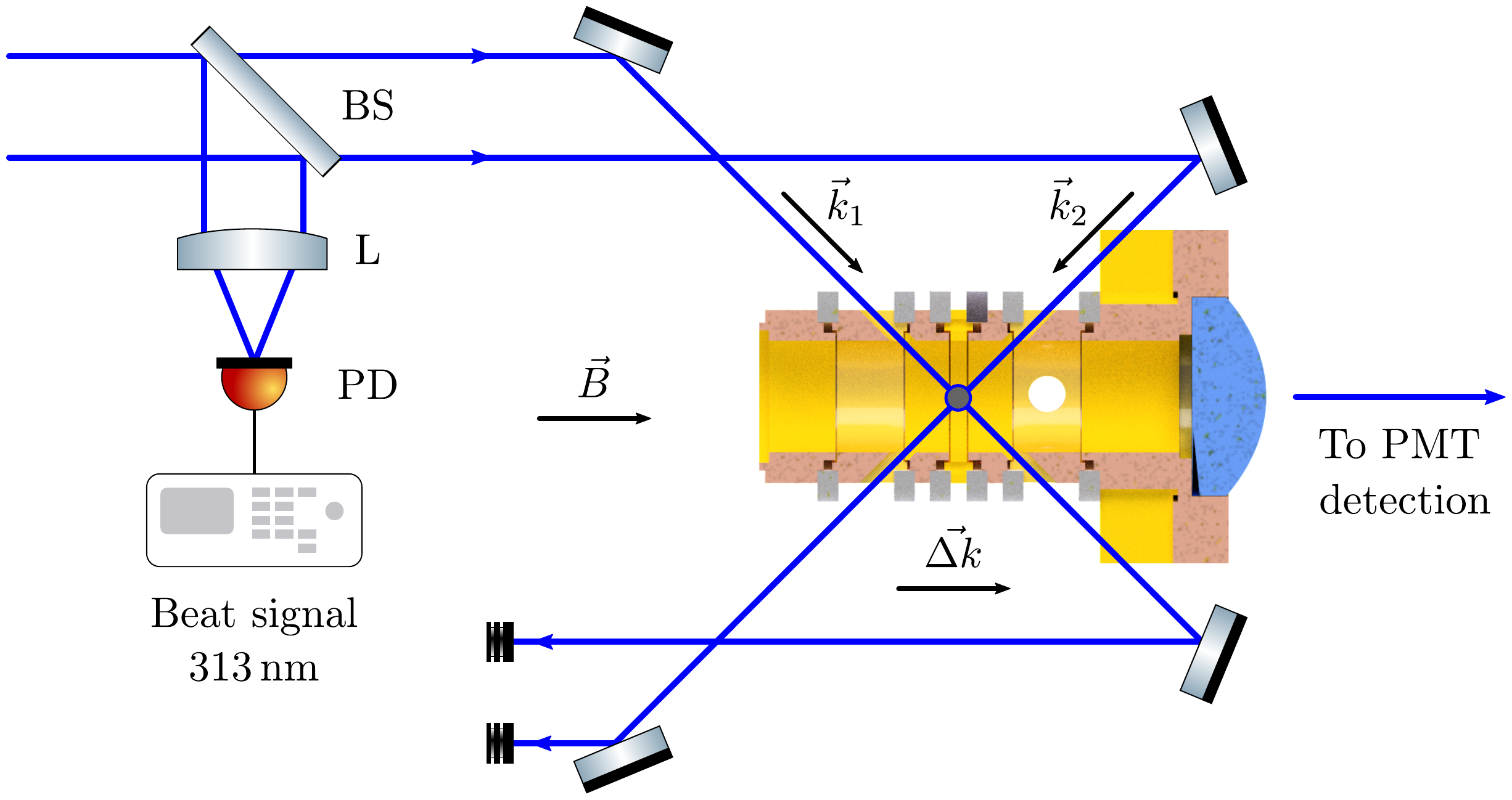}
	\caption{Simplified schematic of the experimental setup for driving stimulated Raman transitions. The orientation of the laser beams with respect to the Penning trap is chosen such that an effective wave vector difference along the magnetic field lines and thus the axial direction is obtained. To obtain a beat signal comparable to the one at the ion cloud position, part of both laser beams is split off with a beam splitter and focused onto a photodiode before sending the beams to the trap. BS: beam splitter, L: lens, PD: photodiode.}
	\label{fig:experimentalsetup}
\end{figure}

As a first demonstration, we verify the phase lock with an offset frequency of 139\,GHz at 313\,nm by performing a thermometry measurement with a \BePlus ion crystal confined in a cryogenic Penning trap. The apparatus used for this measurement is described in detail in Ref.~\cite{niemann_cryogenic_2019}. The ions are trapped in a cylindrical 5-pole trap with an inner diameter of 9\,mm and holes for optical access with laser beams under angles of $45^\circ$ with respect to the trap axis (shown in Fig.~\ref{fig:experimentalsetup}). The angles are chosen such that crossed laser beams can be introduced from the same side of the trap to form a right angle and align their wave vector difference $\vec{\Delta k}=\vec{k}_1-\vec{k}_2$ along the trap axis and magnetic field direction.

We use a trap bias of $V=-20\,\mathrm{V}$ on the ring electrode giving rise to an axial trap frequency of $\omega_z/2\pi=436\,\mathrm{kHz}$ and prepare an ion crystal with an estimated ion number of  about 50 in the trap. The ion number is thereby estimated from fluorescence measurements with a photomultiplier tube (PMT) and 2D images of the ions taken along the axial direction with an EMCCD camera, which we compare with corresponding measurements and images of single ions and few ion crystals, where the ion number can be determined from discrete jumps in the fluorescence level \cite{niemann_cryogenic_2019}.

\begin{figure}[b]
	\centering
	\includegraphics[width=0.6\columnwidth]{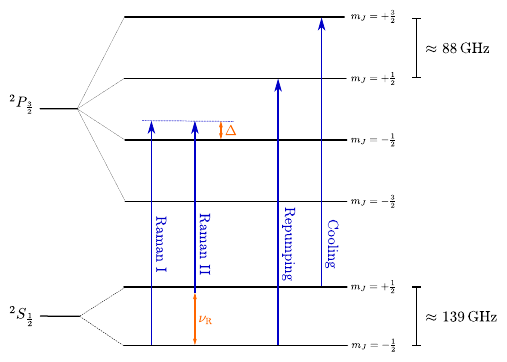}
	\caption{Energy level scheme and relevant transitions for stimulated Raman transitions. The $^2P_{1/2}$ manifold is not shown for clarity and only levels with $m_I=3/2$ are depicted.}
	\label{fig:levelscheme}
\end{figure}

A simplified energy level scheme of \BePlus is shown in Fig.~\ref{fig:levelscheme}. The ions are Doppler cooled and initialized in the $^2S_{1/2}\ket{m_I=3/2,m_J=1/2}$ state with an additional pair of lasers at 313\,nm, which are almost overlapped and cross the trap center with a $45^\circ$ angle relative to the trap axis in the horizontal plane. One of this lasers is tuned about 10\,MHz below the cooling transition ($^2S_{1/2}\ket{m_I=3/2,m_J=1/2}$  $\leftrightarrow$ $^2P_{3/2}\ket{m_I=3/2,m_J=3/2}$) and offset from the trap center in the vertical direction to enable intensity gradient cooling \cite{itano_laser_1982}. The other laser is resonant with the repumping transition ($^2S_{1/2}\ket{m_I=3/2,m_J=-1/2}$ $\leftrightarrow$ $^2P_{3/2}\ket{m_I=3/2,m_J=1/2}$) and roughly aligned to the trap center. As the positions of the laser beams have no absolute reference to our vacuum system and are not actively stabilized relative to the trap, the determination of the offsets from the ion cloud is not straightforward. For the initial alignment we monitor the fluorescence level and spatial distribution of the ion cloud with the EMCCD camera and adjust the laser beam alignment with a piezo motorized focusing lens until a stable signal is observed. Afterwards we regularly optimize the laser beam positions to compensate for drifts relative to the trap and use the results of the thermometry measurements described below to decide whether realignment is necessary.

The two Raman laser beams are detuned by $\Delta\approx20\,\mathrm{GHz}$ from the $^2S_{1/2}\ket{m_I=3/2,m_J=-1/2}$ $\leftrightarrow$ $^2P_{3/2}\ket{m_I=3/2, m_J=-1/2}$ transition (Raman laser I) and the  $^2S_{1/2}\ket{m_I=3/2,m_J=+1/2}$ $\leftrightarrow$  $^2P_{3/2}\ket{m_I=3/2, m_J=-1/2}$ transition (Raman laser II), where Raman laser I is prepared with p-polarization and Raman laser II with right circular polarization. While Raman laser II is overlapped with the repumping laser beam, Raman laser I is aligned to a different trajectory, such that it crossed the other laser beams with an angle $90^\circ$ at the center of the trap (see Fig.~\ref{fig:experimentalsetup}). As the polarization overlap in the reference frame of the ions is higher for Raman laser II, this leads to an asymmetry in the required intensity levels. In the following, we prepare Raman laser I with a beam waist of $145\,\mu\mathrm{m}$ and a power of 3.2\,mW and Raman laser beam II with a beam waist of $120\,\mu\mathrm{m}$ and a power of 1.2\,mW. Due to the alignment of the wavevector difference along the magnetic field, the resulting interaction of the Raman laser beams is limited to the axial motion of the ions and the axial temperature can be retrieved with minimal contributions from the radial motions.

\begin{figure}[h]
	\centering
	\includegraphics[width=0.5\columnwidth]{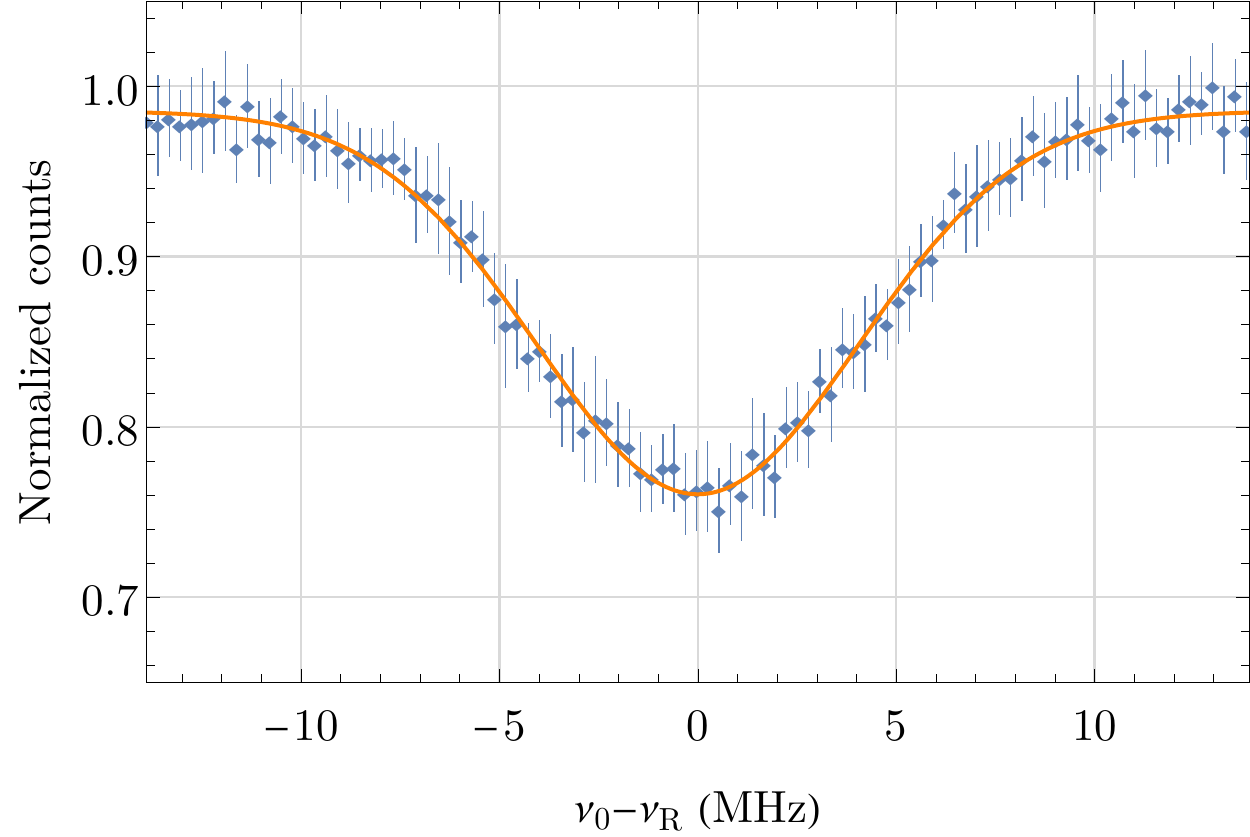}
	\caption{Raman resonance measurement with crossed laser beams. The normalized counts are plotted as a function of the difference of the relative offset frequency between both Raman lasers $\nu_\textnormal{R}$ and the central resonance frequency $\nu_0$. The error bars reflect the standard error (1$\sigma$) of the individually scanned resonance spectra. The solid orange line is a fit of a Gaussian profile to the data. During the measurements we observe a slow irregular drift of the cooling laser beam position. This drift changes the offset of the laser beam relative to the trap center and affects the ion crystal's rotation frequency as well as its size and shape, which in turn affects the overall fluorescence level. While the measured uncertainty is clearly affected by this systematic effect, the goodness of the fit is not. See text for more information.}
	\label{fig:ramanresonance}
\end{figure}

For each data point of the thermometry measurement, we prepare the ion cloud with simultaneous pulses of the cooling and repumping lasers with a length of 75\,ms. Afterwards we shine in the phase-locked laser beams for a time of $500\,\mu\mathrm{s}$ to drive the Raman transition. This transfers part of the ions' population into the $^2S_{1/2}\ket{m_I=3/2,m_J=-1/2}$ state. We then detect by applying the cooling laser again for 12\,ms and collect the fluorescence along the axial direction with the PMT. We repeat the measurement for different offset frequencies between the Raman lasers introduced by a double pass AOM at 313\,nm. An averaged result of 20 individually scanned resonance spectra is shown in Fig.~\ref{fig:ramanresonance}.

 The high number of axial normal modes  prevents the resolution of sidebands and the resulting envelope corresponds to a thermally broadened spectrum \cite{mavadia_optical_2014, wineland_laser_1979}. Performing a Gaussian fit we retrieve a FWHM of $\nu_\textnormal{D}=(9.60\pm0.27)\,\mathrm{MHz}$. Using 
 \begin{equation}
 T=\frac{m\lambda^2\nu_\textnormal{D}^2}{8\ln 2\,k_\textnormal{B}}
 \end{equation}
for a Doppler-broadened spectrum with the beryllium mass $m$, laser wavelength $\lambda$ and Boltzmann constant $k_\textnormal{B}$, we convert this to an axial temperature of $T=(1.77\pm0.10)\,\mathrm{mK}$. This value is a factor of $3.5$ larger that the expected Doppler cooling limit, which can have several possible reasons. On the one hand, a deviation of the Raman laser beam alignment from the ideal one and a residual wavevector difference in the radial direction may lead to a contribution of the radial motion to the measured temperature. In this case, drifts of the cooling laser parameters such as the offset position may contribute to an increased temperature as well. On the other hand, an influence due to the change of the ion cloud dynamics during the spectroscopy pulse because of the absence of the cooling laser beam cannot be ruled out. The detailed investigation of the currently achieved temperature limit will be subject of future experimental studies.
\section{Conclusion}

We have demonstrated phase locking of two lasers at 313\,nm with a frequency offset of 139\,GHz. The offset was generated by modulation of a fundamental laser at 1552\,nm with 4th-order sidebands at 69.5\,GHz using an EOM and phase locking of one sideband to a second laser at 1552\,nm using a digital PLL. Subsequent SFG and SHG have been used to transfer the resulting frequency offset to the UV domain. We have shown the viability of this technique with thermometry measurements of the axial motion of \BePlus ions confined in a cryogenic Penning trap.
The laser system will be used to investigate trap and transport heating phenomena in the future. It might also be used for temperature measurements in sympathetic cooling schemes in Penning traps \cite{bohman_sympathetic_2018}. Moreover, careful preparation and stable trapping of single \BePlus ions in our trap will enable us to measure optical sideband spectra and to apply sideband cooling, which is an essential prerequisite for sympathetic ground state cooling of single \pbars and quantum logic schemes.

\section*{Acknowledgments}
We would like to thank the Laboratory of Nano and Quantum Engineering (LNQE) of Leibniz Universität Hannover for support. We acknowledge funding by the DFG through CRC 1227 ``DQ-mat'' project B06, the cluster of excellence ``Quantum Frontiers'', RIKEN EEE Pioneering Project Funding and ERC StG ``QLEDS''. 

\section*{References}
\bibliographystyle{iopart-num}
\bibliography{qc}

\end{document}